\documentclass[reprint, amsmath, amssymb, aps, prl]{revtex4-1}
\usepackage{braket}
\usepackage{graphicx}
\usepackage{dcolumn}
\usepackage{bm}
\usepackage{xcolor}
\usepackage[normalem]{ulem}

\newcommand{\up}{\ket{\uparrow}} 
\newcommand{\down}{\ket{\downarrow}} 
\newcommand{\Jzproj}{J_{z,\text{QPN}}} 
\newcommand{\DThetaSQL}{\Delta \theta_{\text{SQL}}} 
\newcommand{\ProjNoise}{\Delta \omega_{\text{QPN}}} 
\newcommand{\ProjNoiseOne}{\Delta \omega_{\text{QPN,meas}}} 

\newcommand{\kFSR}{\delta k} 
\newcommand{\Neff}{N_{\text{eff}}} 
\newcommand{\grms}{g_{\text{rms}}} 
\newcommand{\go}{g_0}  
\newcommand{\CouplingFrac}{\zeta} 
\newcommand{\lambdaProbe}{\lambda_p} 

\newcommand{\Nu}{N_{\uparrow}} 
\newcommand{\Nd}{N_{\downarrow}} 
\newcommand{\Jzp}{J_{zp}} 
\newcommand{\Jzf}{J_{zf}} 

\newcommand{\OmegaDownDip}{\omega^\text{D}_\downarrow } 
\newcommand{\OmegaDownDipPre}{\omega^\text{D}_{\downarrow p} }
\newcommand{\OmegaDownDipFin}{\omega^\text{D}_{\downarrow f}}

\newcommand{\OmegaDown}{\omega_\downarrow } 
\newcommand{\OmegaUp}{\omega_\uparrow } 

\newcommand{\NOscT}{\mathcal{N}(t)} 
\newcommand{\NOscF}{|\tilde{\mathcal{N}}(f)^2|}

\newcommand{\gam}{\gamma}
\newcommand{\sigi}{\hat{\sigma}_{z,i}}
\newcommand{\sig}{\hat{\sigma}_{z}}

\newcommand{\Tavg}[1]{\langle #1 \rangle}
\newcommand{\Travg}[1]{\langle #1 \rangle}
\newcommand{\Tw}{T_{\text{win}}} 
\newcommand{\Td}{T_{\text{diff}}} 

\newcommand{\oai}{\hat{\omega}_{mi}}
\newcommand{\opi}{\hat{\omega}_{pi}}
\newcommand{\ofi}{\hat{\omega}_{fi}}

\newcommand{\op}{\hat{\omega}_{p}}
\newcommand{\of}{\hat{\omega}_{f}}
\newcommand{\gai}{g_{mi}}
\newcommand{\gpi}{g_{pi}}

\newcommand{\eai}{\eta_{mi}}
\newcommand{\eaiprime}{\eta_{m'i}}
\newcommand{\epi}{\eta_{pi}}
\newcommand{\efi}{\eta_{fi}}

\newcommand{\ea}{\eta_{m}}
\newcommand{\eaprime}{\eta_{m'}}
\newcommand{\ep}{\eta_{p}}
\newcommand{\ef}{\eta_{f}}
\newcommand{\dc}{\delta_c}
\newcommand{\vardiff}{(\Delta\omega_{\text{diff},i})^2}
\newcommand{\vardifftot}{(\Delta\omega_{\text{diff}})^2}
\newcommand{\diff}{\hat{\omega}_{i,\text{diff}}}
\newcommand{\difftot}{\hat{\omega}_{\text{diff}}}
\newcommand{\Wopt}{W_\textrm{opt}}

\newcommand{\Rmin}{R}
\newcommand{\RminOpt}{R_{\textrm{opt}}}
\newcommand{\vardiffQPN}{(\Delta \omega_\textrm{QPN})^2}

\newcommand{\Eefep}{\Travg{\eta_f \eta_p}}

\newcommand{\Eefsq}{\Travg{\eta_f^2}}

\newcommand{\Eepsq}{\Travg{\eta_p^2}}
\newcommand{\cov}{\textrm{Cov}(\eta_f, \eta_p)}
\newcommand{\Varep}{(\Delta \eta_p)^2}
\newcommand{\Varef}{(\Delta \eta_f)^2}

\newif\ifpdfx
\pdfxfalse

\begin{document}

\title{Spatially Homogeneous Entanglement for Matter-Wave Interferometry Created with Time-Averaged Measurements}
%
%
%

%

\author{Kevin C. Cox}
\affiliation{JILA, NIST, and Department of Physics, University of Colorado, 440 UCB, Boulder, CO  80309, USA}
\author{Graham P. Greve}
\affiliation{JILA, NIST, and Department of Physics, University of Colorado, 440 UCB, Boulder, CO  80309, USA}
\author{Baochen Wu}
\affiliation{JILA, NIST, and Department of Physics, University of Colorado, 440 UCB, Boulder, CO  80309, USA}
\author{James K. Thompson}
\affiliation{JILA, NIST, and Department of Physics, University of Colorado, 440 UCB, Boulder, CO  80309, USA}

\date{\today}

\begin{abstract}
We demonstrate a method to generate spatially homogeneous entangled, spin-squeezed states of atoms appropriate for maintaining a large amount of squeezing even after release into the arm of a matter-wave interferometer or other free space quantum sensor.  Using an effective intracavity dipole trap, we allow atoms to move along the cavity axis and time average their coupling to the standing wave used to generate entanglement via collective measurements, demonstrating 11(1)~dB of directly observed spin squeezing.  Our results show that time averaging in collective measurements can greatly reduce the impact of spatially inhomogeneous coupling to the measurement apparatus.

\end{abstract}

\maketitle
Spin-1/2 atoms must project into either ``up'' or ``down'' when measured.  For $N$ unentangled atoms, the independent randomness in this quantum projection fundamentally limits the single-shot phase resolution of any quantum sensor to $\Delta \phi_{\text{SQL}} = 1/\sqrt{N}$~rad, the standard quantum limit (SQL) \cite{a_winelandCriterion}.  Collective measurements of atoms in optical cavities have recently produced some of the most strongly entangled, spin-squeezed states to date, directly improving the phase resolution of a quantum sensor's ``clock hand'' by a factor up to 60-70 (roughly 18~dB) in noise variance below the SQL \cite{Thompson17dB, Kasevich18dB}.  

Spin-squeezed states could be used to improve a wide range of quantum sensors, with today's best atomic clocks \cite{JunsClock,LudlowClock,katori_clock} being particularly promising candidates \cite{PhysRevA.93.023804,PhysRevA.93.021404}.  In this work we focus on preparing spin-squeezed states appropriate for matter-wave atom interferometry with applications including inertial sensing \cite{Bouyer_review}, measurements of gravity and freefall, \cite{PreciseG_Tino, Lorentz_Rasel} and even the search for certain proposed types of dark matter and dark energy \cite{Reidel, Muller_DE}.

A major challenge arises for cavity-based atom interferometry and other applications involving release of spin-squeezed atoms into free space.  The problem is that the probe mode used to perform the collective measurement is a standing wave, but the atoms are trapped in a 1-dimensional lattice defined by a standing wave cavity mode with a significantly different wavelength.  Some atoms will sit in lattice sites positioned near nodes and some near anti-nodes of the entanglement-generating probe light.  As a result, the atoms will contribute to the collective measurement with different strengths.  In this common case, the large degree of squeezing exists only for this specific coupling configuration and would be largely lost after releasing the atoms into the arm of an interferometer, since their final coupling to the cavity mode or other readout detector will be different from the original configuration \cite{VladanInhomoCoupling}.  In contrast, we wish to create spatially homogeneous entanglement, quantified by the amount of observed phase resolution beyond the SQL that one can achieve when every atom couples equally to the final measurement apparatus.

\begin{figure}[h!]
\centering
\includegraphics[width=3.375in]{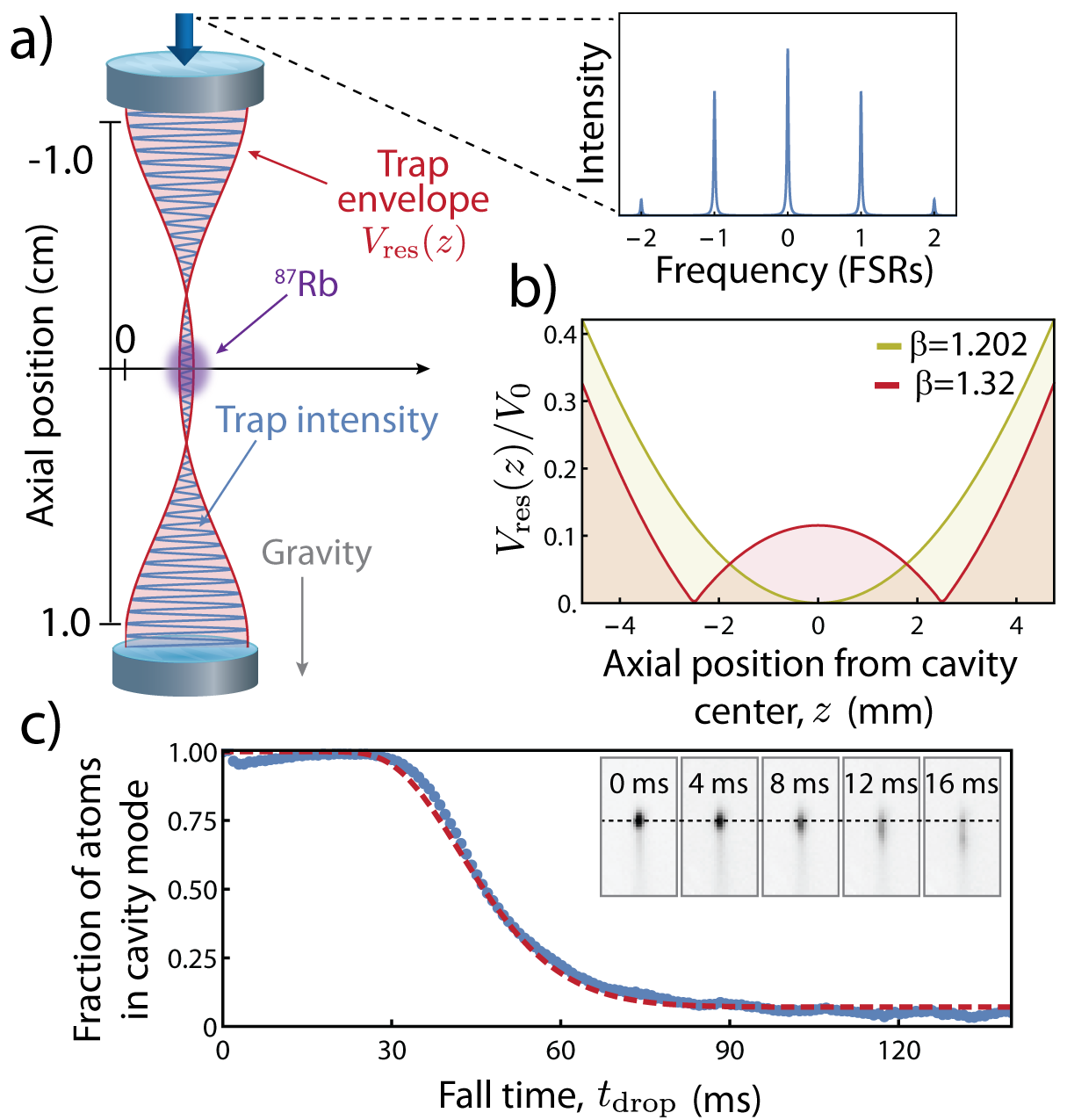}
\caption{ (a) Optical lattice sidebands separated by one free spectral range (FSR) are injected into the cavity to create an axially homogeneous ``dipole'' trap.  
Dipole trap intensity (blue) and its envelope (red) plotted inside of the optical cavity, with exaggerated wavelength $\lambda_l \times 10^3$.  (b) The envelope of the residual lattice potential $V_\text{res}(z)$ normalized to the peak lattice potential depth $V_0$ is plotted near the cavity center, optimized for a minimum at $z=0$ (gold, $\beta = 1.20$) and for the minimal fraction of trapped atoms determined experimentally (red, $\beta = 1.32$). (c) Fraction of atoms remaining in the cavity mode (blue points) vs. fall time, fit to a model (red dash) described in the text. Fluorescence images show the falling atom cloud at various times (inset).  
}
\label{fig1}
\end{figure}
In this Letter, we demonstrate a method to create homogeneous spin-squeezed states in a standing wave optical cavity by allowing the atoms to traverse many wavelengths of the standing wave during each collective measurement.  Atoms experience a time-averaged coupling to the cavity so  that every atom is measured with the same strength, ensuring homogeneous entanglement.  We do this by creating an optical trap with a uniform axial potential, which we refer to as an effective ``dipole trap'' as opposed to the standing-wave ``lattice''.  The dipole trap maintains transverse confinement of the atoms while allowing free movement subject to gravity along the vertical cavity axis.  We demonstrate 11(1) dB of directly observed squeezing via collective measurements in the dipole trap and use fluorescence images and noise scalings to show that the generated squeezing is homogeneously shared among the atoms to a large degree, in principle allowing significant amounts of squeezing for free space or guided matter-wave interferometry. We also discuss the limits placed on entanglement generation with time-averaged measurements.

Homogeneous squeezing can also be obtained using a travelling wave ``ring'' cavity \cite{BouyerRingCavity}, but birefringence must be controlled to maintain the efficacy of utilizing cycling transitions \cite{Chen_SSPRA}. Another appealing approach is to introduce a commensurate lattice \cite{Kasevich18dB, kasevichCommensurate}. This approach  requires special mirror coatings and frequency doubling equipment and doesn't permit guided movement for atom interferometry within the cavity mode. Homogeneous entangled states can also be obtained without using a cavity \cite{OberthalerInt, Appel,a_mitchell_singlet,Oberthaler2014,Chapman12}, but free space experiments have not yet achieved the large amounts of squeezing observed using optical cavities.

In this work, we use the pseudo-spin states defined by the ground hyperfine states of $^{87}$Rb, with $\down \equiv \ket{5^2 S_{1/2}, F = 1, m_F = 1}$ and $\up \equiv \ket{5^2 S_{1/2}, F = 2, m_F = 2}$ split by 6.8~GHz. As in Refs. \onlinecite{Thompson17dB,Chen_SSPRA}, we describe the total pseudo-spin state of $N$ atoms by a collective Bloch vector $\vec{J}$, with spin projections $J_x$, $J_y$, and $J_z$.  The spin projection on a single trial $J_z = \Nu - \frac{N}{2}$ is determined by making a collective measurement of the total number of atoms in the upper spin state $\Nu$.  For an unentangled, coherent spin state (CSS), quantum projection noise (QPN) leads to fluctuations in $J_z$ of size $\Delta \Jzproj = \sqrt{N}/2$.  In this work, $\Delta X$ will refer to the standard deviation of a quantity $X$ as measured over repeated trials of the experiment.

The collective measurement is performed using the experimental apparatus and techniques described in Ref. \onlinecite{Thompson17dB}.  In brief, we trap $^{87}$Rb atoms in the central 2~mm of a 2~cm optical cavity with finesse $F = 2532(80)$.  A cavity mode is tuned $\delta_c = 2 \pi \times 400$~MHz to the blue of the $\up$ to $\ket{e} \equiv \ket{5^2 P_{3/2}, F = 3, m_F = 3}$ transition. The cavity resonance frequency $\omega$ is shifted by an amount depending on the number of atoms in $\up$ due to the dispersive interaction between the atoms and cavity.  The cavity's resonance frequency is measured by probing the cavity in reflection for 40~$\mu$s.   The probing is collective because it is not possible to tell from the single probe mode precisely which atoms are in $\up$.

In a single trial, we apply resonant microwaves to prepare each atom in an equal superposition $(\up+\down)/\sqrt{2}$. We then perform two consecutive measurements of the projection $J_z$, with the two measurement outcomes labeled $\Jzp$ and $\Jzf$, with subscripts denoting pre and final measurement.  The quantum projection noise is common to the two measurements and is removed when we take the difference between the pre and final measurements, yet the atoms nearly completely retain coherence of the quantum phase between $\up$ and $\down$.  This allows one to sense a quantum phase that evolves between the final and premeasurements below the SQL. 

The atoms are initially cooled to approximately 10~$\mu$K and trapped in a far off resonance red detuned optical lattice at $\lambda_l~=~823$~nm (with corresponding wave vector $k_0 = 2 \pi / \lambda_l$). We then convert this standing-wave lattice into an effective dipole trap.  This is achieved by simultaneously driving multiple TEM$_{00}$ longitudinal modes of the cavity near 823~nm. Adjacent longitudinal modes have opposite symmetry with respect to the cavity center.  To lowest order, near the center of the cavity, one mode creates a $\cos^2(k_0 z)$ standing-wave intensity profile while the next mode creates a $\sin^2(k_0 z)$ intensity profile such that the sum of the two standing waves $\cos^2(k_0 z) + \sin^2(k_0 z) =1$  creates a net uniform intensity profile along the cavity axis as shown in Fig.~\ref{fig1}(a).

To drive adjacent longitudinal modes, we phase modulate the lattice light at the cavity free spectral range (FSR), $\text{FSR} = 2 \pi \times 8.1050(5)$~GHz, using a fiber-coupled phase modulator.  The resulting axial component of the potential at distance $z$ from the cavity center can be written $V(z) = V_0[J^2_0(\beta) \cos^2\left( k_0 z\right) + J^2_{-1}(\beta) \sin^2\left( (k_0 + \kFSR_{-1}) z \right) + J^2_{1}(\beta) \sin^2\left( (k_0 + \kFSR_1) z \right) + \dots]$, where $J_n(\beta)$ is the $n$th Bessel function and $\beta$ is the modulation index. $\kFSR_n = n\text{FSR}/c$ is the additional wave vector for the sidebands offset by $n$ cavity free spectral ranges, with speed of light $c$.  Interference terms between sidebands are neglected since they oscillate at 8~GHz. 

Figure~\ref{fig1}(b) shows the depth of the residual standing-wave lattice potential in the dipole trap $V_\text{res}(z)$ as a function of distance from the center of the cavity for two different values of $\beta$.  We find $\beta \approx 1.32$ (overdriving the dipole trap) to be the optimum value for freeing atoms to move.   This is due to a wider minimum of $V_\text{res}(z)$ which overlaps the atomic spatial distribution as well as the fact that overdriving causes the lattice potential wells to be converted into small potential peaks, giving atoms additional potential energy.  

When an atom begins to fall in the dipole trap, the increase in the residual lattice depth is not sufficient to stop the atom from continuing to fall; rather, we expect the atom to be guided by the optical dipole trap until it collides with the lower mirror.  In Fig.~\ref{fig1}(c), we measure the number of atoms in the cavity as a function of freefall time, $t_\text{drop}$, by continuously monitoring the dispersive shift of the cavity resonance frequency.  The data is renormalized to account for background atom loss and is reasonably described by a fit (purple line) which assumes atoms are guided by the net transverse intensity profile of the dipole trap until they are lost when they collide with the lower mirror. For comparison, ballistic expansion out of the cavity mode would occur in only 2~ms were we to simply turn off the optical lattice.  The free fall and guiding are corroborated by fluorescence measurements such as shown in Fig.~\ref{fig1}(c) inset for various $t_\text{drop}$.  Figure \ref{fig1}(c) and fluorescence images indicate that at long times only $5(1)\%$ of the atoms remain trapped in a residual lattice. The majority of the atoms move along the cavity axis, the key for obtaining time-averaged homogeneity in the coupling of the atoms to the standing-wave probe mode.

\begin{figure}[htb]
\includegraphics[width=3.375in]{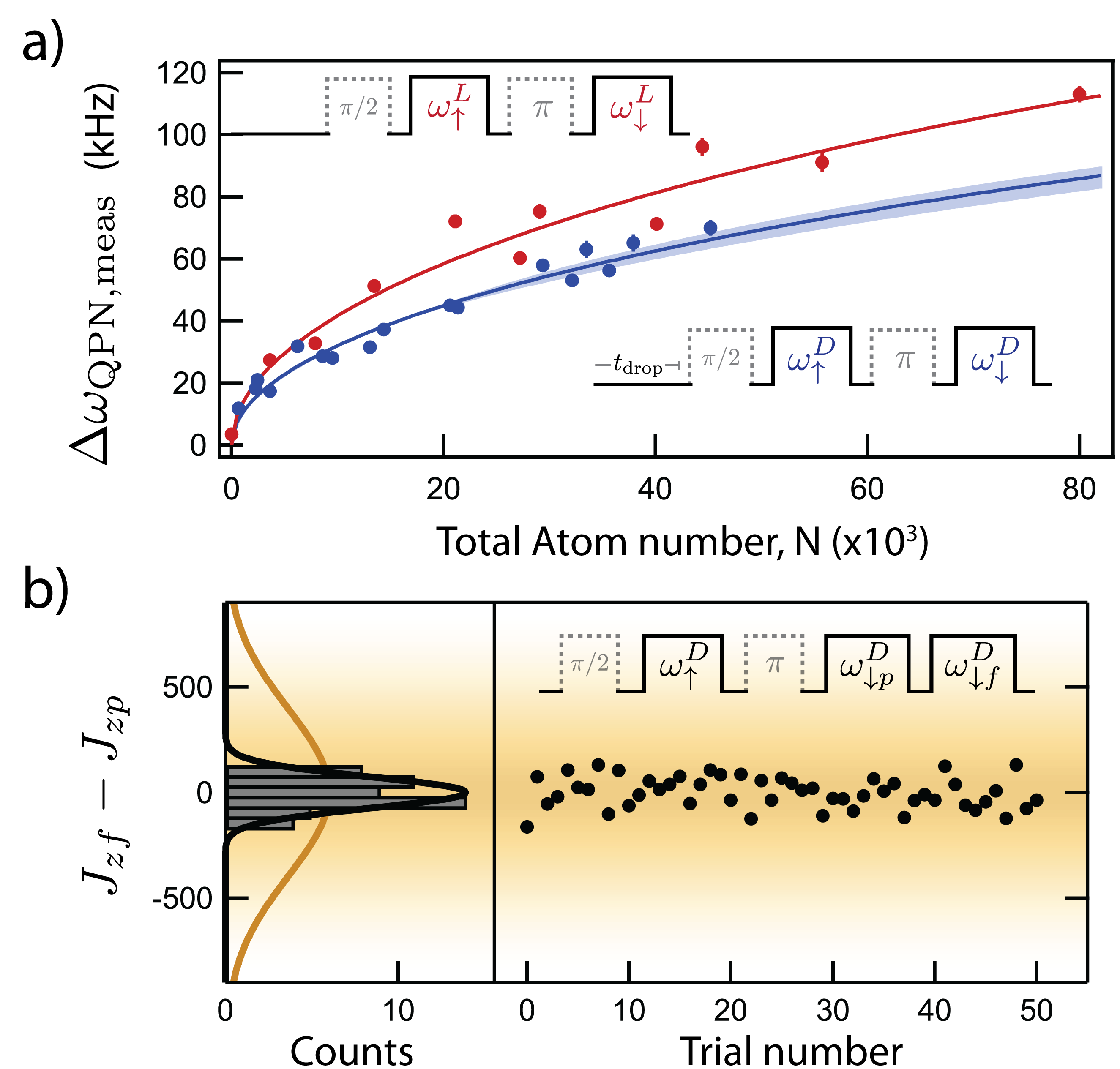}
\caption{(a) Projection noise scaling versus total atom number $N$, measured in the lattice (red points) including a theoretical prediction (red line) and in the dipole trap (blue points) including a fit to infer a coupling fraction $\CouplingFrac$ (blue line, with 68\% confidence interval bands). Sequences are inset. Dashed boxes represent Bloch vector rotations through a given angle using resonant microwaves. Solid boxes represent cavity frequency measurements. (b) Quantum noise reduction in the dipole trap with 6.3(3)$\times 10^5$ atoms. A histogram of $\Jzf - \Jzp$ (black data points) shows a standard deviation 13.9(6)~dB below projection noise $\Delta \Jzproj = 397$ atoms (gold line and shaded distribution).  The measurement sequence is inset.
}
\label{fig2}
\end{figure}

For a fixed total atom number, we expect the projection noise induced fluctuations in the cavity resonance frequency $\ProjNoise$ to be smaller in the dipole trap than in the lattice.  While the total dispersive shift is the same in both cases, in the lattice the dominant contribution is from the subset of atoms situated near antinodes of the probe.  These atoms have a Jaynes-Cummings coupling parameter $g_i$ near the maximum value $g_0 = 2 \pi \times 0.519(5)$~MHz and provide stronger than average fluctuations.  In the ideal time-averaged situation, on the other hand, the full ensemble only couples with the rms coupling strength $g_{rms} = g_0/\sqrt{2}$, actually leading to weaker cavity frequency fluctuations.  To quantify the level of homogeneous coupling, we define a model where fractionally, $\CouplingFrac$ of the atoms release into the dipole trap and are assumed to have perfectly homogeneous coupling.  $1-\CouplingFrac$ of the atoms remain fixed in position and maintain their original coupling. In this model, the projection noise induced fluctuations in the cavity resonance frequency can be written $\ProjNoise = \grms^2 \sqrt{N(3-\CouplingFrac)}/\sqrt{8(g_0^2 N+\delta_c^2)}$ \cite{Note1}.

We observe this change in the projection noise scaling between the lattice and dipole trap by performing the measurement sequences of Fig.~\ref{fig2}(a) in the lattice (red, superscript L) and in the dipole trap (blue, superscript D) versus the total atom number in the cavity $N$.  The $\OmegaUp$ and $\OmegaDown$ windows represent the outcome of a measurement of the cavity resonance frequency, sensitive to $\Nu$ or $\Nd$ respectively, and we plot the observed projection noise fluctuations $\ProjNoiseOne = \Delta(\OmegaUp-\OmegaDown)$ in either the lattice or the dipole trap.  A small amount of technical noise that does not have the proper scaling with atom number has been subtracted out of this data.  
The lattice data is used as a calibration of $g_0$ with the theoretical scaling plotted in red.  The dipole trap data is fit to the model $2\times \ProjNoise$ (since the measurement sequence includes two anti-correlated windows, $\OmegaUp$ and $\OmegaDown$) with $\CouplingFrac$ as a free parameter. We fit $\CouplingFrac = 1.0(2)$, consistent with our expectation of 95\% from the data in Fig. \ref{fig1}(c).  

By consecutively performing a pre and final measurement $\OmegaDownDip$, labeled $\OmegaDownDipPre$ and $\OmegaDownDipFin$ we can show a large degree of spin noise reduction below QPN and correspondingly demonstrate the creation of entangled, spin-squeezed states in the dipole trap.  We measure spin squeezing using the Wineland criterion for phase enhancement relative to the SQL, $\left( \Delta \theta/\DThetaSQL \right)^2 \equiv S = R / C^2$ \cite{Wineland1992,Thompson17dB}.  The observed spin noise reduction normalized to the quantum projection noise level is $R = (\Delta (\Jzf - \Jzp) / \Delta \Jzproj)^2 < 1$. Squeezing or enhanced phase resolution also requires the additional demonstration of retained coherence, or Bloch vector length, often referred to as ``contrast'', $C \equiv 2 |\vec{J}|/N$.

The measurement sequence is shown in the inset of Fig. \ref{fig2}(b) and is the same as that of Ref.~\cite{Thompson17dB}.  We use $t_\text{drop}~=~13$~ms, which accelerates the atoms enough to average over approximately 13 cycles of the probe standing wave during the 40~$\mu$s measurement window.  Figure \ref{fig2}(b) shows noise in measurements of $\OmegaDownDipFin-\OmegaDownDipPre$ in the dipole trap with total atom number $N = 630(30)\times 10^3$~atoms.  Experimental parameters $g_{rms}$, $\delta_c$, and $N$ are used to scale between cavity frequency measurements and $J_z$, $\partial \omega / \partial J_z = g_{rms}^2 / \sqrt{4 g_{rms}^2 \Nu +\delta_c^2}$.  The data is collated into a histogram on the left, showing a standard deviation $13.9(6)$~dB less than the projection noise level shown in yellow.  The remaining contrast after the premeasurement was independently measured, $C=$0.70(5).  Together with the noise reduction, this yields a directly observed phase resolution, or spin squeezing, of $S=1/13(3)$ or  $-11(1)$~dB below the SQL  

\begin{figure}
\centering
\includegraphics[width=3.375in]{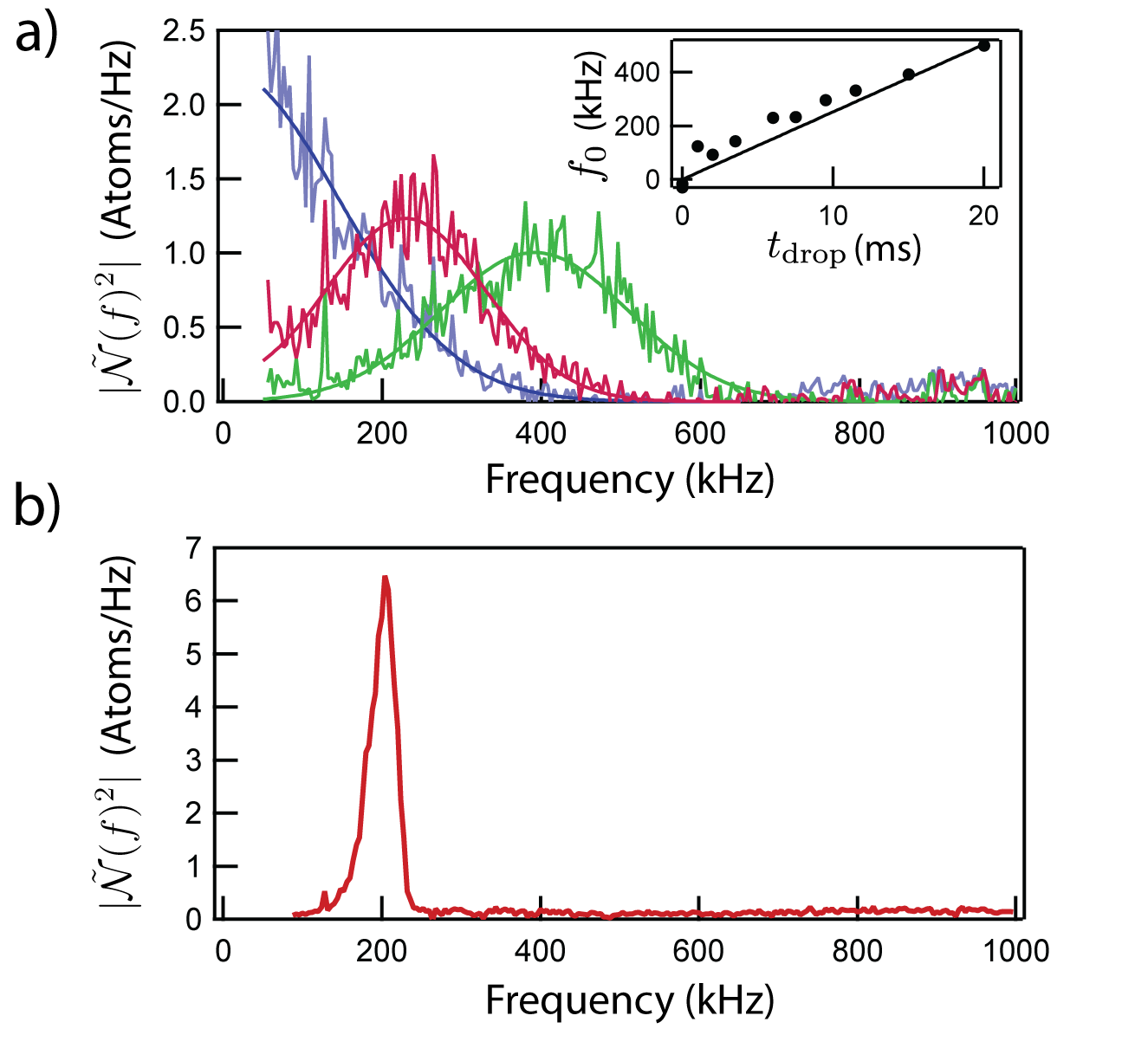}
\caption{(a) Power spectra showing coupling oscillations for fall times of 1 ms (blue), 7.5 ms (red) and 15 ms (green) with their respective fits. (inset) Center frequency $f_0$ of the fitted Boltzmann distribution for various fall times (points) compared to a freefall prediction line of $f_0 = 2 a t / \lambdaProbe$ (line), see text for definitions.  (b) Power spectrum showing coupling oscillations at the trap frequency when atoms are trapped in the optical lattice.}
\label{fig3}
\end{figure}

When the cavity frequency is measured in a 40~$\mu$s window using the dipole trap, oscillations in the signal are observed, indicating the atomic motion over the probe standing wave.  Specifically, we measure the number of atoms $\Nu$ that are coupled to the cavity as a function of time by applying a scale factor to convert cavity frequency to atom number. We refer to this rescaled time signal as $\NOscT = \sum_i^{N_{\uparrow}} g_i^2(t) / \grms^2$.  We observe noise in the atom's coupling in the frequency domain, which can be used to infer the distribution of atoms' coupling oscillation frequencies. Most of the coupling oscillations average away, since the oscillation of each atom occurs with a random phase.  However, the residual uncancelled coupling oscillations are observed in $\NOscT$ such that the squared Fourier transform of the time signal, $\NOscF$ has units of Atoms/Hz and is closely related to the atomic velocity distribution.

Figure \ref{fig3}(a) shows $\NOscF$, recorded using 2~ms of data and taking the average power spectrum of time traces from approximately 65 trials.  The data was taken after 1~ms (blue), 7.5~ms (red), and 15~ms (green) of freefall time after release into the dipole trap. Each power spectrum is fit to an appropriately folded 1D Boltzmann distribution that accounts for the inability to distinguish between upwards and downwards velocities.  The fit center frequency $f_0$ is plotted as a function of the freefall time $t$ in the inset of Fig. \ref{fig3}(a).  The result is consistent, particularly at long times, with the simple prediction, $f_0 =a \,t / (\lambdaProbe/2)$, where $a=9.81$~m/s$^2$ is the acceleration due to gravity. The widths of the distributions are consistent with Boltzmann distributions giving final axial temperatures of 25~$\mu$K.  To contrast, Fig.~\ref{fig3}(b) shows $\NOscF$ for atoms in the lattice. Instead of a large thermal distribution, a narrow distribution is observed at the lattice trap frequency, about 200~kHz. 

In summary, we infer that we have created a spatially homogeneous squeezed state from the combined observations of Figs.~1-3.  First, we observe release of 95(1)\% of the atoms (Fig. \ref{fig1}) at a sufficient velocity (Fig. \ref{fig3}) to ensure, on average, 13 averaging cycles of the probe standing wave during a 40~$\mu$s collective measurement.  In Fig. \ref{fig2}(a) we also confirm the transformation to homogenous coupling by the change in scaling of projection noise fluctuations of the cavity.  The demonstration of 11(1)~dB of observed squeezing in the homogeneous configuration proves our ability to create a large amount of entanglement in this highly time-averaged scheme.

We investigate the limits to squeezing using our time-averaged scheme in the Supplemental Material \cite{Note1}.  Since the ac signals in Fig.~\ref{fig3} yield additional information about the spin state of each velocity component of the atomic ensemble, time averaging will fundamentally limiting the squeezing to order $S\propto \frac{1}{q N}$, where $q$ is the total quantum efficiency of the experiment.  For $q \sim 1$, this is close to the Heisenberg limit.  The more relevant limitation for our system is imperfect averaging of the probe standing wave.  For 40~$\mu$s measurements after a 13~ms drop time and for a 25~$\mu$K ensemble, we estimate that the observed noise reduction should be limited to 15~dB below QPN, which we believe to be a primary limitation to our observed spin noise reduction of 13.9(6)~dB.  In the future, this limit could be improved using longer measurement windows to average over more cycles of the probe standing wave during each measurement. 

To realize a free space matter-wave interferometer, atoms could be prepared in the cavity for the entanglement generating premeasurement, then released into free space for an interferometry sequence.  The final measurement could be performed fluorescence detection.  The 5\% of atoms with non-uniform coupling during the premeasurement would lead to an additional noise floor, not observed in this work, of 13~dB below the SQL.  Additionally, inhomogeneity from radial motion will lead to another additional noise floor of approximately 10~dB below the SQL \cite{Thompson17dB}.  Notably, this radial motion would also equally affect systems using ring cavities, commensurate lattices, or other axial averaging techniques.

Another possibility is to perform guided interferometry inside the cavity mode.  Here the pre and final measurements would both be performed with collective cavity measurements.  In this case, the noise from the 5\% of atoms remaining trapped and radial motion largely cancels at short times.  The 11(1)~dB of squeezing observed in this work would in principle fully translate to this type of interferometer.  In addition to the possibility of using entangled states, performing the final readout via a cavity measurement may allow for reduced technical noise, higher bandwidth, cleaner optical modes, and power buildup for Raman transitions \cite{HolgerCavityInterf15}. 

Similarly, higher order transverse modes, atom-chip technologies \cite{ DanaBECChipInterferometer, AtomChipSagnac}, or tailored potentials \cite{Reichel_GuidedInt,klitzing_Painted} might be combined with the cavity measurement technique presented here to create new varieties of matter-wave Sagnac interferometers and other inertial sensors.  The real-time observation of mechanical motion also opens the path to stochastic cooling schemes based on measurement and feedback \cite{Vladan_CoolingFB} with applications to more complex systems such as molecules, which can be challenging to laser cool using conventional Doppler cooling methods.

We gratefully acknowledge support from NIST, DARPA QuASAR, ARO, and NSF PFC.  This  material  is  based  upon  work  supported by the National Science Foundation under Grant Number 1125844 Physics Frontier Center.

\nocite{Vuletic_QND, Squeezing_Bohnet_2014,LSM10,CBS11,CavPhotonFluct}
\bibliography{main}

\pagebreak
\widetext
\begin{center}
\textbf{\large Supplemental Material:  Spatially Homogeneous Entanglement for Matter-Wave Interferometry Created with Time-Averaged Measurements}
\end{center}
\setcounter{equation}{0}
\setcounter{figure}{0}
\setcounter{table}{0}
\setcounter{page}{1}
\makeatletter
\renewcommand{\theequation}{S\arabic{equation}}
\renewcommand{\thefigure}{S\arabic{figure}}
\renewcommand{\bibnumfmt}[1]{[S#1]}
\renewcommand{\citenumfont}[1]{S#1}

\section{Introduction}
As shown in the Main Text, time averaging the probe standing wave during collective cavity measurements can be used to create homogeneous squeezing within an atomic ensemble in a cavity.  However, the technique introduces limitations to spin squeezing in addition to the fundamental limits calculated in Ref. \onlinecite{Chen_SSPRA} set by  finite collective cooperativity $N \mathcal{C}$, finite quantum efficiency for detecting the probe light $q$, and the probability of the probe light inducing a spin flip $p$.  Specifically, the best achievable spin squeezing is  $S = \mathrm{Max}\left[e/(q N \mathcal{C}), \sqrt{8 p/(qN\mathcal{C})}\right]$ where the first limit ($e/(qNC)$) is due to wave-function collapse from free space scattering and the second limit  ($\sqrt{8 p/(qN\mathcal{C})}$) is due to additional noise from Raman spin flips.  Here, we calculate the additional limits to $S$ imposed by time averaging the spatially inhomogeneous coupling.

First, we establish the language for describing the limits on spin-noise reduction $R$ when the coupling coefficient of each atom differs between the pre  and final measurement windows, with results closely matching those of Hu \textit{et al.} \cite{VladanInhomoCoupling}.  We apply this formalism to determine the best achievable spin-noise reduction, and therefore best spin squeezing for two concrete scenarios relevant to matter-wave interferometry with squeezed states.  First, we consider the scenario in which we perform the premeasurement of the spin state with the atoms trapped in an incommensurate 1D intracavity lattice and then perform the final measurement with uniform coupling.  Next, we consider the case of incomplete cycle averaging over the standing wave probe mode if the atoms are allowed to move along the cavity axis during the pre and final measurements.

Next, we calculate two sources of inhomogeneous broadening, or dephasing, that arise from time averaging.  This dephasing leads to a loss of contrast or signal. The first source is fundamental quantum back-action that arises from the atoms sampling the photon shot noise of the intracavity probe light at different times, or equivalently, spectral frequencies. The second source arises from imperfect cycle averaging of the probe coupling.  We show that these two limiting effects do not significantly impact current experiments, and can in principle allow spatially homogeneous squeezing near the Heisenberg limit.

\section{Noise Reduction Limits with Inhomogeneous Coupling}
\subsection{Noise between the final and premeasurement}
To find the optimum spin-noise reduction with inhomogeneous coupling, we begin with an expression for a measurement of atoms in a cavity where we assume each atom $i$ couples to the optical cavity with a factor $\eai(t)$ in the pre ($m = p$) or final ($m = f$) measurement.  Concretely, we can write $\eai(t) = \gai^2(t)/\dc$ when we are in the dispersive regime of cavity measurements, where $\dc$, the optical cavity detuning from the atomic transition, is much greater than $\Omega =2 g_\text{rms} \sqrt{N}$, the collectively enhanced coupling rate between the atoms and the cavity, or vacuum Rabi splitting \cite{Chen_SSPRA}.  $\gai^2(t)$ is the $i$th atom's time dependent Jaynes-Cummings coupling parameter.   We consider the average outcome of the measurement by taking the time average of $\eai(t)$, denoted by dropping the time dependence $(t)$.  As a reminder, $\grms^2$ is the average of $g_i^2$ over the atomic ensemble.  Note however, that by changing the units and scaling of $\eai$, the expressions we derive can be generalized for any type of population measurement such as fluorescence detection.  The operator $\oai$ that measures the time-averaged cavity frequency shift from the $i$th atom is then

\begin{align}
\label{eq1}
\oai = \left(\sigi' + \gam \right) \eai.
\end{align}
Here,  $\sigi' =(1-\gam)(\, \ket{\uparrow_i} \bra{\uparrow_i} - \ket{\downarrow_i} \bra{\downarrow_i}\,)$ is the Pauli spin operator $\sigi$ for the $i$th atom, rescaled by $(1-\gam)$.  The constant $0\le\gam\le1/2$ is used to account for the fact that the measurement may be sensitive to some linear combination of state populations $\Nu$ and $\Nd$, given by $(\Nu - (1-2\gam)\Nd)$.  We include this factor since a number of spin squeezing experiments have used an optical cavity detuned halfway between the $\up \rightarrow \ket{e}$ and $\down \rightarrow \ket{e}$ transitions \cite{Kasevich18dB, Vuletic_QND, Appel}.  In such cases, the dispersive cavity shift is sensitive only to population differences between $\up$ and $\down$ and can be modeled with $\gam = 0$.  However, in this work we set the cavity resonance frequency near the $\up \rightarrow \ket{e}$ transition (modeled with $\gam = 1/2$), so there is no sensitivity to $\Nd$.  This scheme is advantageous for probing on a cycling transition $p=0$, as mentioned in the introduction \cite{Thompson17dB,Chen_SSPRA,Squeezing_Bohnet_2014}.  Additionally, even when the cavity only couples to atoms in $\up$, a $\pi$-pulse can be used to swap the state populations and measure $\Nu$ and $\Nd$ sequentially, again achieving a differential measurement with $\gam = 0$. However, this method is only valid if the atoms' couplings to the measurement do not change between the $\Nu$ and $\Nd$ measurement.

We wish to cancel the quantum projection noise in the final measurement of all $N$ atoms $\of=\sum_{i=1}^N \ofi$ using a premeasurement $\op=\sum_{i=1}^N \opi$ of the noise value.  However, if the $i$th atom's couplings $\efi$ and $\epi$ differ, its projection noise cannot be exactly cancelled.  In order to optimize the cancellation of the noise, we  construct a weighted difference $\difftot \equiv \of - W \op$, with weight factor $W$.  It is important to note that, since our entanglement-generating collective measurements of $\of$ and $\op$ must not reveal single-particle information, we can only use a single weight factor for the entire ensemble.  Thus, if the change between $\efi$ and $\epi$ is inhomogeneous, that is, different for each atom $i$, there is no value of $W$ that can be chosen to achieve full cancellation of the noise in the final measurement.

To derive the best achievable squeezing limit with inhomogeneous coupling, we will calculate the noise in the weighted difference $\difftot$ and find the optimum value for the weight factor $W$.  First, using Eq.~\ref{eq1}, we calculate the variance $\vardiff$ in  $\diff$ for a single atom.  Next, we will independently sum the noise contributions from each atom to calculate the total variance $\vardifftot$.  We neglect all sources of noise in the system except the quantum projection noise and fluctuations in the couplings $\eai$.  Other realistic noise sources such as photon shot noise and laser frequency noise are neglected in order to calculate the squeezing limit just from inhomogeneous coupling.  Importantly, the optimum weight factor $W$ for cancelling quantum noise and coupling noise is likely to not be the optimum for cancelling the other technical noise sources such as laser frequency noise.  In fact, in our experiments optimum squeezing was always observed with $W = 1$.

The noise variance in $\diff$ is calculated using
\begin{align}
\vardiff \equiv \Travg{(\ofi -W \opi)^2} - \Travg{(\ofi -W \opi)}^2,
\end{align}
\noindent where we denote the average over many independent experimental trials with the $\Travg{...}$ notation.  This trial average will simultaneously evaluate the quantum fluctuations of the spin projection operator $\sigi'$ as well as possible fluctuations in the time-averaged couplings $\eai$.  In calculating averages, we will reasonably assume that fluctuations in the couplings are uncorrelated with fluctuations in the spin projection operator, leading to

\begin{align}
\vardiff &= \Travg{(\sigi' + \gam)^2}\Travg{(\efi - W \epi)^2} \\
&- \Travg{\sigi' + \gam}^2\Travg{\efi - W \epi}^2.
\end{align}
We now limit the discussion to the relevant case of atoms in an equal superposition of $\up$ and $\down$ such that  $\Travg{\sigi'}=0$.  However, we leave the expectation value $\Travg{\sigi'^2}$ unevaluated.  This helps to illuminate different contributions to $\vardiff$ and also allows for the possibility of atoms being in different entangled or mixed states.
\begin{align}
&\vardiff = \Travg{\sigi'^2}\Big [\Travg{\efi^2 + W^2\epi^2-2W\efi \epi}\Big] \label{wquant}\\
& + \gam^2 \Big[ (\Delta \efi)^2 + W^2(\Delta \epi)^2 -2 W \text{Cov}(\efi ,\epi) \Big]\label{wclas}
\end{align}
where $\text{Cov}(X,Y) = \Travg{XY}-\Travg{X}\Travg{Y}$ is the covariance between X and Y.  As a reminder, the covariance quantifies the correlation of the fluctuations of $X$ and $Y$.  If $X$ and $Y$ are uncorrelated, then $\text{Cov}(X,Y) =0$.  If $X$ and $Y$ are perfectly correlated, then $\text{Cov}(X,Y) =\Delta X \Delta Y$ where $\Delta X$ is the standard deviation in the quantity $X$.  There are two contributions to $\vardiff$.   The first (line \ref{wquant}) is the term arising from uncancelled quantum projection noise and will be nonzero if the couplings, $W \epi$ and $\efi$, are not equal to one another. This term captures the fundamental problem of inhomogeneous coupling to the collective measurement.

The second contribution proportional to $\gamma$ (line \ref{wclas}) results from trial-to-trial noise in the couplings $\efi$ and $\epi$ and is fully classical, since it has no contribution from projection noise, that is, $\Travg{\sigi'^2}$. The physical origin of this term can be thought of as trial-to-trial noise in the scale factor relating an observed cavity frequency shift to an estimated population of atoms in spin up or down.  When $\gamma=0$, the cavity frequency shifts are proportional to $N_\uparrow-N_\downarrow$ which is on average zero for the case considered here. As a result, any noise in the scale factor contributes no additional noise.  However, if $\gamma=1/2$, then the cavity frequency shifts are proportional to $N_\uparrow$ which for the case considered here is approximately $N/2$.  In this case, classical scale factor noise can easily contribute at or above the quantum noise level. 

As an example, the second term (line \ref{wclas}) is important when the atoms are trapped in the intracavity lattice for the premeasurement and then released to into an optical dipole trap for the final measurement.  An atom in the optical lattice sees fluctuations with a range of 100\% from trial to trial in the standing wave coupling $\epi$.  This leads to classical noise in the premeasurement that is of the same order as quantum projection noise.  The quantum (line \ref{wquant}) and classical (line \ref{wclas}) contribution to $\vardiff$ are approximately equally.  The classical noise could, in principle, be removed if we knew every atom's couplings $\efi$ and $\epi$ on each trial, but we do not.  Instead, we can only measure the average couplings $\Travg{\efi}$ and $\Travg{\epi}$ by linking a cavity shift, averaged over many trials, to a change in the state population $\Nu$.

The second relevant example that is sensitive to both quantum and classical noise in $\difftot$ is that of our time-averaged scheme. Atoms will have different initial positions and velocities from trial to trial, leading to classical noise between $\efi$ and $\epi$.  We theoretically predict the contribution to our observed squeezing from this effect on Page 5.

The total noise variance $\vardifftot$ is found by summing the noise contribution from each atom, assuming that they are uncorrelated:

\begin{align}
\vardifftot = \sum_{i=1}^N \vardiff.\label{test8}
\end{align}
Every second order moment (e.g. $\Travg{\eai^2}$) contributing to this sum can in principle be different, but we make the often-valid assumption that the particle labels for every atom are interchangeable.  In this case, the trial average $\Travg{...}$ can equivalently be viewed as an average over all of the atoms in the ensemble on a single trial.   In either picture, each term in  Eq.~\ref{test8} is equal so that
\begin{align}
\vardifftot &= N \vardiff \label{vartot}.
\end{align}
Since we have assumed that expectation values are the same for  all atoms $i$, we will drop the indeces in following expressions such that $\Travg{\eai \eaiprime} \equiv \Travg{\ea \eaprime}$, $\Travg{\eai} \equiv \Travg{\ea}$, and $\Travg{\sigi'^2} \equiv \Travg{\sig'^2}$.

\subsection{Quantum projection noise}

To calculate the quantum projection noise (QPN) limit to the noise difference $\vardifftot$, we can consider the case in Eq.~\ref{wquant}-\ref{wclas} when $W=0$ and only take the resulting noise from the quantum term of line \ref{wquant},
\begin{align}
\vardiffQPN = N \Travg{\sig'^2} \Travg{\ef^2}.\label{testqpn}
\end{align}
Equation \ref{testqpn} can be used to derive the prediction (given in the Main Text) for projection noise fluctuations in the effective dipole trap when fractionally $(1-\CouplingFrac)$ of the atoms remained trapped in the residual lattice.  We assume that the atoms are identically prepared in a pure state with an equal superposition of $\up$ and $\down$, so that $\Travg{\sigi'^2} = 1/4$.  We have set $\gam =1/2$, to model our measurements that are sensitive to $\Nu$.  For the $\CouplingFrac$ atoms in the dipole trap, $\Travg{\ef^2} = \go^4/(4\dc^2)$ where $\go$ is the Jaynes-Cummings coupling parameter at an anti-node of the cavity.  For the $1-\CouplingFrac$ atoms in the residual lattice, $\Travg{\ef^2} = 3 \go^4/(8 \dc^2)$.  Simply adding the projection noise variance of all of the atoms leads to the observed decrease of the projection noise versus $\CouplingFrac$,
\begin{align}
\vardiffQPN(\CouplingFrac) =\frac{N \go^4}{4\dc^2} \left[ \frac{1}{4} \CouplingFrac +  \frac{3}{8}(1-\CouplingFrac)\right].
\end{align}
The equation for the QPN level given in the Main Text is more general in that it does not assume the dispersive limit $\dc \gg 2\sqrt{N}g_\text{rms}$ as we have done here.

\subsection{Optimum spin-noise reduction}

The observable spin-noise reduction relative to the quantum noise in the final measurement is given by $\Rmin=\vardifftot/\vardiffQPN$. Here we consider atoms independently prepared in a pure state with equal superposition of $\up$ and $\down$, such that $\Travg{\sig'^2}=1/(1-\gam)^2$. We then find,
\begin{align}
& \Rmin = \frac{\Travg{\ef^2 + W^2\ep^2-2W\ef\ep}}{\Eefsq} \label{rquant}\\
&+\frac{ \gam'^2}{\Eefsq} \Big[ (\Delta \ef)^2 + W^2(\Delta \ep)^2 -2 W \text{Cov}(\ef ,\ep)\Big] \label{rclas} 
\end{align}

\noindent where we define $\gam' = \gam/(1-\gam)$.  As a reminder, $\gam = \gam' = 0$ represents a measurement of $\Nu- \Nd$ and $\gam = 1/2$, $\gam' = 1$ represents a measurement of only $\Nu$.  Similar to Eq. \ref{wquant}-\ref{wclas}, this result possesses a quantum (line \ref{rquant}) and classical (line \ref{rclas}) contribution.  The classical term is given by the variances, or fluctuations of measurement strengths while the quantum term is given by the squared magnitude of the pre and final measurement coupling strengths.  Additionally, if $\gam' = 0$ the classical term vanishes, since the measurement signal will be centered around zero.  If $\gam' = 1$, the two components can be of the same size.  This shows that engineering measurements sensitive to $\Nu-\Nd$ can be advantageous to measuring $\Nu$.

The optimum weight factor that minimizes $\Rmin$ is
\begin{align}
\Wopt = \frac{\gam'^2 \text{Cov}(\ef ,\ep) +  \Travg{\ef \ep}}{\Eepsq + \gam'^2 (\Delta \ep)^2},
\end{align}
with optimized noise reduction $\RminOpt$,
\begin{align}
&\RminOpt   = 1+\frac{ \gam'^2 \Varef }{\Eefsq} \label{Ropt1} \\
&-\frac{ \left[ \Eefep + \gam'^2 \cov   \right]^2}{\Eefsq \Eepsq +  \gam'^2 \Eefsq \Varep} \label{Ropt2}.
\end{align}
This result is the best noise reduction possible when the final and premeasurement have different coupling strengths.  Line~\ref{Ropt1} gives the projection noise and classical noise in the final measurement.  Line~\ref{Ropt2} represents the optimum cancellation of the final measurement's noise provided by the optimally weighted premeasurement.   In the case that only population differences ($\Nu - \Nd$) are mesured, such that $\gamma' = 0$, $\RminOpt$ simplifies to
\begin{align}
\RminOpt = 1-\frac{\Eefep^2}{\Eefsq \Eepsq},
\end{align}
the ratio of the second order moments of the coupling strengths.

\subsection{Fundamental limits for specific cases}\label{examps}
We now apply the previous results to two important cases.  First, we show that spin-squeezed states created in an optical lattice with incommensurate coupling to the probe standing wave do not lead to significant noise reduction below the SQL after they are launched into a homogeneous environment such as free space or our time-averaged optical dipole trap.  Second, we derive the limit to spin-noise reduction in our time-averaged scheme due to imperfect time averaging of the probe standing wave.
\subsubsection{Inhomogeneous pre versus homogeneous final}
For applications such as atom interferometry, it is interesting to consider the case in which the atoms uniformly couple to the final measurement (homogeneous coupling), but during the premeasurement the atoms are held in an optical lattice that has an incommensurate wavelength with the probe standing wave (inhomogeneous coupling.)  

We take the final coupling to be the same for all atoms $\ef = \go^2/2\dc$. This represents the case when the final measurement is performed with the atoms moving along the cavity axis such that they perfectly time average away spatially inhomogeneous coupling to the probe standing wave.  This will also capture the physics of similar such measurement scenarios including ring cavities, commensurate lattice/probe standing waves and spatially homogeneous fluorescence detection.

In our experiment, the lattice $\lambda_l= 823$~nm and probe $\lambda_p=780$~nm wavelengths are higly incommensurate.  The two standing wave antinodes go from fully aligned to misaligned to fully realigned in roughly $7.5~\mu$m, a much shorter length than the characteristic 1~mm range of lattice sites into which atoms are loaded.  To describe this scenario, we take the premeasurement coupling of the $i$thatom to depend on its fixed position $x_i$ on a single trial as $\epi = (\go^2/\dc) \sin^2(\Phi_i)$, where the phase of the coupling is $\Phi_i= 2 \pi x_i / \lambda_p$.   We further assume that the atoms independently and randomly load into different lattice sites from one trial to the next such that the $i$thatom uniformly samples coupling phases $\Phi_i= 0$ to $2\pi$ from trial to trial.  However, we re-emphasize that to model a premeasurement performed with the atoms trapped in the lattice, on a single trial the coupling phase of atom $i$ does not vary in time.    One then finds second order moments $\Travg{\ep^2} = 3 \go^4/8 \dc^2$ and $\Travg{\ef^2}=\Travg{\ep \ef} =\go^4/4\dc^2 = \grms^4/\dc^2$, and variances $\Varep= \Travg{\ep^2}- \Travg{\ep}^2 =  \go^4/8 \dc^2$, $\Varef=0$ and $\text{Cov}(\ef ,\ep) =0.$

The best achievable spin-noise reduction for this case is $\RminOpt = 1/3$ ~$(1/2)$ or -4.8~dB (-3~dB) when we consider the two different cases, $\gamma'=0~(1)$. This shows that squeezed states created by a premeasurement in an incommensurate standing wave lattice \cite{Thompson17dB} cannot provide  significant entanglement enhancement for free space sensors.  Specifically, this degree of spin-noise reduction is far worse than the $R \approx -18$~dB of spin-noise reduction achieved in experiments in which the atoms are trapped for both the pre and final measurement \cite{Kasevich18dB,Thompson17dB}. 

This optimal spin-noise reduction is achieved using the optimum weight factor $\Wopt= 2/3$~$(1/2)$.  For comparison, if the relative weight factor is not optimized, but simply set to $W=1$, one finds that $\RminOpt =1/2$ $( 1)$ or -3~dB (0~dB).

For the case $\gamma'=0$, it is interesting to note that in previous  squeezing work with incommensurate standing waves \cite{Vuletic_QND,LSM10,CBS11,Squeezing_Bohnet_2014,Thompson17dB}, an effective atom number $\Neff$ was defined in relationship to the total atom number $N$ as $\Neff= 2 N/3 = \Wopt N$.  The above results provide a nice physical interpretation of this effective atom number.  The premeasurement can in principle perfectly measure the spin noise of $\Neff$ of the total atoms $N$, here 2/3 of all the atoms.  The spin noise of the $\Neff$ atoms can then be perfectly cancelled from the final measurement of $N$ atoms.  The remaining spin noise of the ``unmeasured" atoms $N-\Neff$, here 1/3 of the total atoms, cannot be canceled at all such that the best achievable spin-noise reduction is $\RminOpt=(N-\Neff)/N=1/3$.

\subsubsection{Imperfectly time-averaged pre and final measurements}

As another example, consider the case in the Main Text, where both the pre and final measurements are performed by time averaging the probe coupling as the $i$thatom moves along the cavity axis at velocity $v_i$.  The atom moves from anti-node to anti-node of the probe at frequency $f_i= 2 v_i/\lambda_p$, a frequency we call the coupling oscillation frequency.  The time dependent coupling can then be written as $\eai(t) = (\go^2/\dc) \sin^2( \pi f_i t + \phi_{mi})$, where $\phi_{mi}$ sets the coupling at $t=0$.  We assume both the pre and final measurements last for a time $\Tw$, and that the pre and final measurements start at $t=0$ and $\Td$ $(\Td \ge\Tw)$ respectively. We take the projection noise level to be that for a perfectly time-averaged scenario in which each atom moves exactly an integer number of cycles of the standing wave: $\vardiffQPN = N \go^4/16 \dc^2$.  Under the same assumption, we set the weight factor $W=1$. The spin-noise reduction, averaging over the normalized thermal velocity distribution of atoms $P(f_i)$, is
\begin{align}
\label{fulleq}
\Rmin = \int_{-\infty}^{\infty} P(f_i) \frac{4 \sin ^2( \pi f_i \Td) \sin ^2( \pi f_i \Tw)}{f_i^2 \pi^2 \Tw^2} \,d f_i.
\end{align}
To gain some insight, consider an example when the atomic distribution $P(f_i)$ is Gaussian with mean $f_0$ and standard deviation $\Delta f$. In the limit that $\Td,\,\Tw \gg 1/\Delta f$ and $f_0 \gg \Delta f$, the terms $\sin^2(\pi f_i \Td)$ and $\sin^2(\pi f_i \Tw)$ in Eq.~\ref{fulleq} will oscillate rapidly with $f_i$ and so can be replaced in the integrand by their average 1/2. The resulting spin-noise reduction is then $R = 1/(\pi N_{\text{osc}})^2$ where $ N_{\text{osc}} = f_0 \Tw$ is the number of cycles averaged by an atom at coupling oscillation frequency $f_0$.

We estimate the maximum possible spin-noise reduction expected for the conditions used for spin squeezing in the Main Text, for which the above simplifying approximations are not valid.  To do this, we keep the full expression in Eq.~\ref{fulleq}, set $\Td = \Tw = 40~\mu$s, and use the directly measured distribution of coupling oscillation frequencies shown in Fig~3(a) of the main text to obtain an experimentally measured probability distribution $P(f_i)$.  We find a limit from imperfect averaging $R \approx -15$~dB.  We believe this is one of the primary limits to the observed spin-noise reduction $R= -13.9(6)$~dB.  However, we expect that this limit can be improved to beyond 20~dB by allowing the atoms to fall for longer or by using longer measurement windows $\Tw$, changes that are difficult to implement with current technical constraints of the experiment but that could be straightforward to implement in the future.

\section{Dephasing from Inhomogeneous Back-Action}

In the Main Text, the atoms are allowed to move along the cavity axis in order to achieve time averaging of the inhomogeneous coupling to the standing wave probe mode. As atoms traverse the probe standing wave, they produce fast oscillations in the measured cavity frequency shift in addition to the desired average pre and final measurement signals that are used to resolve phases below the standard quantum limit.  This power spectrum of the oscillations in the cavity shift is shown in Fig.~3 of the Main Text with the spread in frequency components reflecting the thermal spread in atomic velocities.  These oscillating signals yield information about the spin state of atoms moving at a particular velocity and therefore must cause some degree of additional quantum collapse or back-action, which may limit the amount of squeezing.  A full treatment of this effect is very difficult since it likely moves the total wave function away from the restricted fully symmetric Hilbert space of $N$ dimensions and toward the full $2^N$ dimensional Hilbert space.  However, we attempt to derive an estimate for the scale of the deleterious back-action using a classical back-action model driven by quantum noise on the optical probing field.  We estimate that the time-averaging scheme, at worst, only provides an additional squeezing limit near the Heisenberg limit $\Delta \theta_H^2=1/N^2$~rad$^2$.  This is far from being a relevant limit for the best current experiments that achieve enhanced phase resolutions of approximately $\Delta \theta^2 \approx 10^4/N^2$~rad$^2$ with approximately $N = 10^6$ atoms \cite{Kasevich18dB,Thompson17dB}.

\begin{figure}
\centering
\includegraphics[width=3.375in]{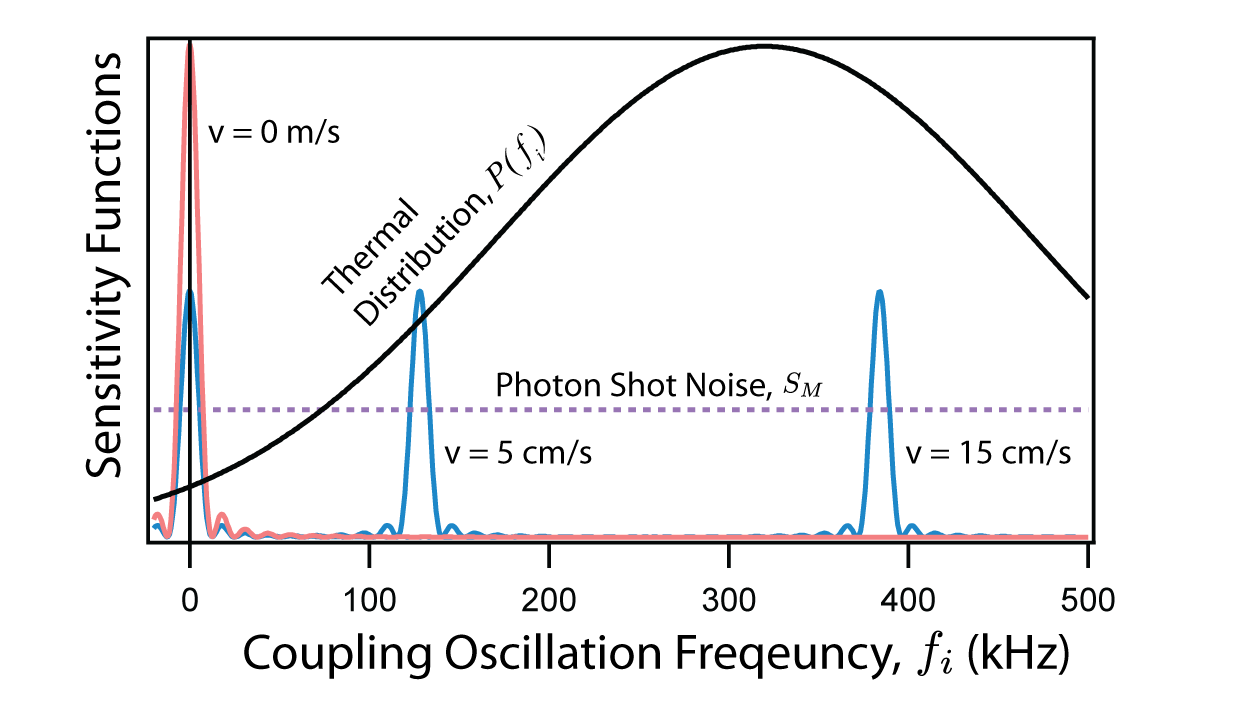}
\caption{Sensitivity of atoms at different velocities to photon shot noise.  A moving atom couples to the probe mode with a transfer function (blue) with sensitivity at DC and at a frequency $f_i$ corresponding to its velocity, shown for an atom with velocity 5~cm/s and 15~cm/s.  Stationary atoms only couple at DC (red).  The distribution of oscillation frequencies is given by the Boltzmann distribution $P(f_i)$ (black).  Atoms at different frequencies sample photon shot noise (PSN, purple dash) at different frequencies leading to dephasing that can limit squeezing with time averaging.}
\label{fig1SM}
\end{figure}

One explanation of the quantum back-action is that the probe light causes a differential AC Stark shift between the spin states such that $\ket{\uparrow} + \ket{\downarrow} \rightarrow \ket{\uparrow} + e^{\imath\psi_i}\ket{\downarrow}$, equivalent to the $i$th atom's Bloch vector changing its azimuthal angle by $\psi_i$. Photon shot noise (PSN) of the probing beam causes a noisy, unknown contribution to the phase shift with rms fluctuation $\Delta \psi$ that is equivalent to the observed quantum back-action in the azimuthal angle.  The PSN level can be plotted (dotted purple line in Fig.~\ref{fig1SM}) in frequency space as a power spectral density of photon number fluctuations in the cavity $S_M = 4 M_c/\kappa$ \cite{CavPhotonFluct}, valid for frequencies much less than the cavity linewidth $\kappa$, where $M_c$ is the average number of photons in the cavity mode and $\kappa$ is the cavity linewidth.  Stationary atoms sample this PSN in a frequency window centered at zero frequency, with characteristic bandwidth $1/\Tw$ (the exact sensitivity function shown in red in Fig.~\ref{fig1SM})  However, if an atom moves along the cavity axis, its coupling to the probe field during the premeasurement oscillates as $g_{pi}^2(t) = g_0^2 \sin^2(\pi f_i t + \phi_{pi})$, and it will sample the PSN with a modified transfer function, sensitive at DC (due to the time-averaged component of $g_{pi}^2(t)$) as well as a component oscillating at $f_i$ (shown for two different velocities or $f_i$ in blue in Fig \ref{fig1SM}).  In the time domain picture, this is equivalent to the intra-cavity photon number fluctuating on a time scale $1/\kappa$.  Atoms at different velocities and initial positions sample these fluctuations differently.  As a reminder, we only consider the back-action of the premeasurement (p) because loss of coherence during or after the final measurement does not affect the desired sensitivity for detecting a phase that is applied between the pre and final measurement. 

Given a thermal velocity distribution of atoms, the distribution of coupling oscillation frequencies $f_i$ will be given by a Gaussian probability distribution $P(f_i)$ with standard deviation  $2 \pi \times \Delta f = 4 \pi  \sqrt{k_b \Tw/(m \lambda_p^2)}$ where $k_b$ is Boltzmann's constant, and $m$ is the mass of $^{87}$Rb.\color{black}  The number of sub-ensembles that receive uncorrelated back-action will be of order $N_e \sim \Delta f \times \Tw$.   If  $N_e\gg 1$, the total Bloch vector length will be reduced, leading to a loss of contrast $C = e^{-\Delta \psi ^2/2}$. We can write the rms phase shift about the mean due to back-action $\Delta \psi$ as 
\begin{align}
\Delta \psi = \alpha \sqrt{M_{r}},
\end{align}
where $M_{r}$ is the total number of probe photons reflected from the cavity input mirror during a measurement, assuming a loss-less, single-ended cavity.  On resonance, the intra-cavity photon number $M_c$ can be related to $M_r$ by $M_r =  M_c (\Tw \kappa)/4 $.   The constant $\alpha = 4 \grms^2 / (\kappa \dc)$ characterizes the average azimuthal phase shift to an atom's Bloch vector per reflected photon.

Using the Heisenberg uncertainty relation for a collective spin state, $\Delta \theta \Delta \psi \ge 1/N$, the PSN limited spin-noise reduction, ignoring all other noise sources, can be shown to be written
\begin{align}
\label{eq10}
R = \frac{1}{\alpha^2 M_r q N  }.
\end{align}
In real experiments, the squeezing is fundamentally limited by contrast loss from free space scattering or diffusion of the Bloch vector due to Raman transitions \cite{Chen_SSPRA}.  However, here we neglect these limits to squeezing and instead focus on the squeezing limit solely due to the contrast loss $C$ from inhomogeneous back-action due to time averaging.  In this case, the total squeezing as a function of $M_r$ becomes
\begin{align}
S =R/C^2= \frac{e^{\alpha^2 M_r}}{\alpha^2 M_r q N  }.
\end{align}
which has an optimum value
\begin{align}
S_\textrm{opt} = \frac{e}{N q}.
\end{align}
$S_\textrm{opt}$ represents an estimate of the quantum limit to squeezing with our time averaging scheme.  However, Equations 20 and 21 show that, for $q$ near 1, the squeezing is only affected by time averaging near the Heisenberg limit.

The physical mechanism and scaling of the squeezing limit (Eq. 21) is quite similar to regular quantum back-action, where every atom receives an identical random phase with an rms magnitude of $ \psi_\text{rms}$ due to photon shot noise in the measurement.  In this standard case, since the phase shift is the same for all atoms, the Bloch vector retains its rms extent $J^2 \equiv \Tavg{\hat{J}_x^2} + \Tavg{\hat{J}_y^2} \approx N^2/4$.  The difference in the time-averaged case is that each subensemble receives a random phase causing a decrease in the collective Bloch vector extent $J^2 < N^2/4$.  In this case, the intrinsic phase resolution of the ensemble is lowered, but only when one approaches the Heisenberg limit.

\section{Dephasing from Imperfect Time Averaging}

In addition to quantum back-action driven dephasing, there can also be classical dephasing:  each atom receives a different average AC Stark shift due to imperfect cycle averaging of the probe standing wave. In realistic experiments, this classical dephasing will usually be much larger than the quantum dephasing described in the previous section. 

The phase shift on a single atom can be written, for a single premeasurement window,
\begin{align}
\psi_i = \int_{0}^{\Tw} dt \frac{\gpi^2(t) M_c}{\delta_c}
\end{align}
where $M_c$ is the average intracavity photon number, taken to be constant for this calculation.  To estimate the dephasing due to  classical imperfections in the time averaging, we calculate the standard deviation in $\psi_i$, $\Delta \psi$, over the atomic distribution .  The coupling for the $i$th atom is $\gpi^2(t) = \go^2 \sin^2(\pi f_i t + \phi_{pi})$, with a normalized probability distribution for $f_i$ and $\phi_{pi}$, denoted $P(f_i,\phi_{pi})$.
\begin{align}
\Delta \psi^2  & =\int_{0}^{2 \pi} d\phi_{pi} \int_{-\infty}^{\infty} df_i P(f_i,\phi_{pi})  \psi_i^2 - \\
&  \left(\int_{0}^{2 \pi} d\phi_{pi} \int_{-\infty}^{\infty} df_i P(f_i, \phi_{pi})  \psi_i \right)^2
\end{align}
Assuming the phase $\phi_{pi}$ of the coupling oscillations is random for each atom, the result can be simplified to
\begin{align}
\Delta \psi^2 = \int_{-\infty}^{\infty} df_i P(f_i) \frac{1}{2} \left( \frac{\go^2 M_c}{2   \pi f_i \delta_c} \right) ^2 \sin^2 (\pi f_i \Tw),
\end{align}
reducing the contrast after a single premeasurement by $e^{-\Delta \psi^2/2}$.  Similar to the uncancelled noise reduction of Eq. \ref{fulleq}, this result has the interpretation of being due to the  phase shift from the final, uncancelled non-integer fraction of an atom's coupling oscillation.  However, the result will be slightly modified by the use of a spin-echo pulse for the premeasurement, as in Fig.~2 (b) of the Main Text, but since the phase of each atom's coupling oscillation changes for each window based on it's velocity, significant spin-echo cancellation of this dephasing is not expected.  For our system we estimate that classical dephasing from imperfect time averaging leads to a small contrast loss of less than $1~$dB at the optimal squeezing and could be improved to arbitrary levels with better averaging by increasing the number of periods of oscillation in a measurement window while holding the total number of incident photons in the window fixed.

\bibliographystyle{apsrev4-1}
\bibliography{main}

\end{document}